\documentclass[twocolumn,amsmath,amssymb,aps,secnumarabic,%
    nofootinbib,superscriptaddress]{revtex4}
\usepackage{times,mathptmx,amsthm,pstricks,pst-plot,pst-node}
\newtheorem{theorem}{Theorem}[section]
\newtheorem{lemma}[theorem]{Lemma}
\theoremstyle{remark}

\newcommand{\C}{\mathbb{C}}
\newcommand{\Z}{\mathbb{Z}}
\renewcommand{\tensor}{\otimes}
\renewcommand{\hat}{\widehat}
\renewcommand{\tilde}{\widetilde}
\DeclareMathOperator{\Tr}{Tr}
\DeclareMathOperator{\im}{im}

\newcommand{\cA}{\mathcal{A}}
\newcommand{\cB}{\mathcal{B}}
\newcommand{\cC}{\mathcal{C}}
\newcommand{\cD}{\mathcal{D}}
\newcommand{\cE}{\mathcal{E}}
\newcommand{\cF}{\mathcal{F}}
\newcommand{\cM}{\mathcal{M}}
\newcommand{\cP}{\mathcal{P}}
\newcommand{\cX}{\mathcal{X}}
\newcommand{\cY}{\mathcal{Y}}

\newcommand{\hcA}{\hat{\cA}}
\newcommand{\hcB}{\hat{\cB}}
\newcommand{\typ}{{\mathrm{typ}}}
\renewcommand{\d}{\mathrm{d}}
\newcommand{\tO}{\tilde{O}}

\newcommand{\eps}{\epsilon}
\newcommand{\pe}{\preccurlyeq}
\newcommand{\injects}{\hookrightarrow}
\newcommand{\binjects}{\stackrel{b}{\hookrightarrow}}
\newcommand{\nimplies}{{\hspace{1ex}\not\hspace{-1ex}\implies}}

\newcommand{\ie}{\textit{i.e.}}

\newcommand{\eatline}{\vspace{-\baselineskip}}

\psset{linewidth=.5pt,dash=3pt 3pt,doublesep=.05,arrowsize=2pt 3}
\newgray{gray80}{.8}

\newenvironment{fullfigure}[2]
    {\begin{figure}[htb]\def\ffa{#1}\def\ffb{#2}}
    {\caption{\ffb.}\label{\ffa}\end{figure}}

\newcommand{\thm}[1]{Theorem~\ref{#1}}
\renewcommand{\sec}[1]{Section~\ref{#1}}
\newcommand{\lem}[1]{Lemma~\ref{#1}}
\newcommand{\eq}[2]{\begin{equation}\label{#1}#2\end{equation}}
\newcommand{\fig}[1]{Figure~\ref{#1}}

\begin{document}
\title{The capacity of hybrid quantum memory}

\author{Greg Kuperberg}
\email{greg@math.ucdavis.edu}
\thanks{Supported by NSF grant DMS \#0072342}
\affiliation{UC Davis}

\begin{abstract} The general stable quantum memory unit is a hybrid consisting
of a classical digit with a quantum digit (qudit) assigned to each classical
state. The shape of the memory is the vector of sizes of these qudits, which
may differ. We determine when $N$ copies of a quantum memory $\cA$ embed in
$N(1+o(1))$ copies of another quantum memory $\cB$.  This relationship captures
the notion that $\cB$ is as at least as useful as $\cA$ for all purposes in the
bulk limit.  We show that the embeddings exist if and only if for all $p \ge
1$, the $p$-norm of the shape of $\cA$ does not exceed the $p$-norm of the
shape of $\cB$.  The log of the $p$-norm of the shape of $\cA$ can be
interpreted as the maximum of $S(\rho) + H(\rho)/p$ (quantum entropy plus
discounted classical entropy) taken over all mixed states $\rho$ on $\cA$.  We
also establish a noiseless coding theorem that justifies these entropies. The
noiseless coding theorem and the bulk embedding theorem together say that
either $\cA$ blindly bulk-encodes into $\cB$ with perfect fidelity, or $\cA$
admits a state that does not visibly bulk-encode into $\cB$ with high
fidelity.

In conclusion, the utility of a hybrid quantum memory is determined by its
simultaneous capacity for classical and quantum entropy, which is not a finite
list of numbers, but rather a convex region in the classical-quantum entropy
plane.
\end{abstract}
\maketitle

\section{Introduction}
\label{s:intro}

Many questions in quantum information theory involve both quantum and classical
information.  The usual computational model for such dual information is
independent quantum and classical memory.  The measurement algebra of a
combined memory consisting of an $a$-state qudit and a $b$-state classical
digit is
$$ \cM_a \tensor \C^b = \bigoplus_{k=1}^b \cM_a,$$
where $\cM_a$ is the set of $a \times a$ matrices.  But this is not the most
general possible hybrid of classical and quantum memory.  Rather the
measurement algebra $\cA$ of a finite memory could be any direct sum of matrix
algebras of possibly different dimensions:
$$\cA \cong \bigoplus_{k=1}^n \cM_{\lambda_k}.$$
The partition (\ie, non-negative integral vector) $\lambda = \lambda(\cA)$ is a
list of the dimensions of the matrix algebras called the \emph{shape} of the
memory $\cA$. \sec{s:memory} discusses why this is a reasonably general quantum
memory model.

For example, the simplest hybrid memory is a \emph{hybrid trit}, with shape
$(2,1)$. It consists of matrices of the form
$$\left(\begin{array}{cc|c} * & * & 0  \\ * & * & 0 \\ \hline 0 & 0 & *
\end{array}\right).$$
This memory models a three-state system in which one state is observed by the
environment but the other two remain coherent relative to each other.  It is
easy to compare the capacity of the hybrid trit to any other quantum memory: 
It is between a qubit and a qutrit, more than a classical trit, less than any
larger memory that contains a qubit, and neither more nor less than a classical
digit with at least 4 states.

It turns out that there is more than one notion by which one memory unit has
more capacity than another. (Atypically, all such notions are equivalent for
the hybrid trit.)  The strictest relevant relationship between memories is
given by algebra embeddings.  If $\cA \injects \cB$ is an algebra embedding
(which need not be unit-preserving, or unital), then the memory $\cB$ can
simulate the memory $\cA$.  In other language, an algebra embedding is a blind,
perfect-fidelity decoding.  \sec{s:memory} also explains that although other
blind, perfect-fidelity encodings are possible, any such encoding can be
replaced by an algebra embedding. As \sec{s:bin} explains, the question of
whether $\cA$ embeds in $\cB$ is a computable (but NP-hard) bin-packing
problem.

In this article we will consider a more relaxed comparison, namely whether many
copies of $\cA$ embed in slightly more copies of $\cB$. More precisely we say
that $\cA$ \emph{bulk-embeds} in $\cB$, or $\cA \binjects \cB$, if for every
rational $\eps > 0$, there exists an $N$ such that
$$\cA^{\tensor N} \injects \cB^{\tensor N(1+\eps)}.$$
If $\cA$ bulk-embeds in $\cB$, there is no reason to pay more for $\cA$ than
$\cB$ when buying large quantities of the two memories with equal performance.
Our first main result is a characterization of when $\cA$ bulk-embeds in $\cB$:

\begin{theorem} If $\cA$ and $\cB$ are two hybrid memories,
then $\cA \binjects \cB$ if and only if
$$||\lambda(\cA)||_p \le ||\lambda(\cB)||_p$$
for all $p \in [1,\infty]$.
\label{th:embed} \end{theorem}

One direction of \thm{th:embed} is straightforward.  The $p$-norm of a
partition $\lambda$ is defined as
$$||\lambda||_p = \biggl(\sum_k \lambda_k^p\biggr)^{1/p}.$$
It is easy to check that the $p$-norm is multiplicative:
$$||\lambda(\cA \tensor \cB)||_p = ||\lambda(\cA)||_p ||\lambda(\cB)||_p$$
for any pair of memories $\cA$ and $\cB$. On the other hand the bin-packing
model implies that if $\cA$ embeds in $\cB$, then
$$||\lambda(\cA)||_p \le ||\lambda(\cB)||_p.$$
It follows that this inequality also holds when $\cA$ bulk-embeds
in $\cB$.  The proof of the other direction of \thm{th:embed}
is the topic of \sec{s:embed}.

The $p$-norm has an interesting information-theoretic interpretation. In
\sec{s:entropy} we will define the classical entropy $H(\rho)$ and the quantum
entropy $S(\rho)$ of a state $\rho$ of a quantum memory $\cA$.  Their
definitions are justified by a capacity estimate, \thm{th:cap},
and by a noiseless coding theorem, \thm{th:code}.

\begin{theorem} Every state $\rho$ of a memory $\cA$
satisfies inequality
$$\frac{H_\cA(\rho)}{p} + S_\cA(\rho) \le \log ||\lambda(\cA)||_p,$$
where $\rho$ has classical entropy $H(\rho)$ and quantum entropy $S_\cA(\rho)$.
For each $p \ge 1$ there exists a $\rho$ that achieves equality. Any
non-negative pair $(H,S)$ satisfying the inequality for all $p$ can be
expressed as
$$(H,S) = (H_\cA(\rho)+t,S_\cA(\rho)-t)$$
for some $\rho$ and some $t \in [0,1]$.
\label{th:cap} \end{theorem}

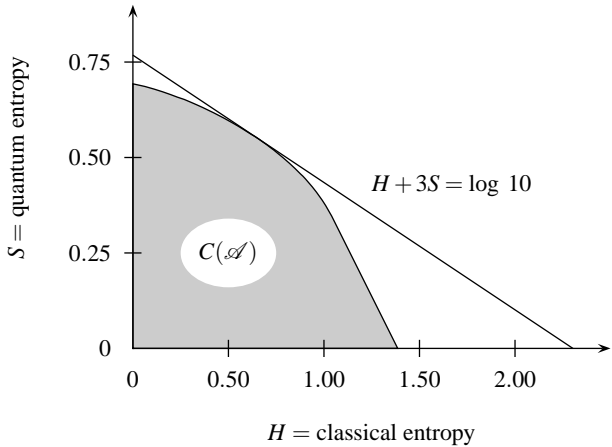
\begin{fullfigure}{f:capacity}{The capacity region of a memory
    $\cA$ with shape $(2,1,1)$, and its $3$-norm bounding line}
\psset{xunit=2.54,yunit=5.08}
\pspicture(-.75,-.3)(2.65,1)
\pspolygon[fillstyle=solid,fillcolor=gray80]
    (0.000,0.693) (0.063,0.686) (0.112,0.679) (0.156,0.672) (0.196,0.665)
    (0.233,0.658) (0.269,0.652) (0.302,0.645) (0.334,0.638) (0.365,0.631)
    (0.394,0.624) (0.423,0.617) (0.450,0.610) (0.477,0.603) (0.502,0.596)
    (0.527,0.589) (0.551,0.582) (0.574,0.575) (0.596,0.568) (0.618,0.561)
    (0.639,0.555) (0.660,0.548) (0.679,0.541) (0.699,0.534) (0.717,0.527)
    (0.736,0.520) (0.753,0.513) (0.770,0.506) (0.787,0.499) (0.803,0.492)
    (0.819,0.485) (0.834,0.478) (0.849,0.471) (0.863,0.464) (0.877,0.457)
    (0.890,0.451) (0.903,0.444) (0.915,0.437) (0.927,0.430) (0.939,0.423)
    (0.950,0.416) (0.961,0.409) (0.971,0.402) (0.981,0.395) (0.991,0.388)
    (1.000,0.381) (1.009,0.374) (1.017,0.367) (1.025,0.360) (1.033,0.353)
    (1.040,0.347) (1.386,0) (0,0) (0,0.693)
\psaxes[Dx=.50,Dy=.25]{<->}(2.5,.9)
\psline(0,0.768)(2.303,0)
\rput(1.25,-.225){$H = $ classical entropy}
\rput{90}(-.6,.5){$S = $ quantum entropy}
\psellipse*[linecolor=white](.5,.25)(.25,.09)
\rput[bl](1.25,.4){$H+3S = \log\ 10$}
\rput(.5,.25){$C(\cA)$}
\endpspicture
\end{fullfigure}

Note that the three most common $p$-norms are also significant for quantum
information theory.  The logarithm of the 1-norm, $\log ||\lambda(\cA)||_1$, is
the purely classical capacity of $\cA$. The logarithm of the $\infty$-norm,
$\log ||\lambda(\cA)||_\infty$, is the purely quantum capacity. And the
logarithm of the $2$-norm,
$$\log ||\lambda(\cA)||_2 = \frac{\log\ \dim \cA}{2},$$
is half of the dense coding capacity of $\cA$.

\thm{th:cap} implies that the set of possible pairs
$$(H_\cA(\rho)+t,S_\cA(\rho)-t),$$
where $0 \le t \le S_\cA(\rho)$, forms a convex capacity region $C(\cA)$ in the
first quadrant of the plane. \fig{f:capacity} shows an example.  The constant
$t$ expresses the fact that quantum entropy can be used classically. Since the
$S$-intercept of the line tangent to $C(\cA)$ with slope $-\frac1p$ is $\log\
||\lambda(\cA)||_p$, another way to state \thm{th:embed} is that memory $\cA$
bulk-embeds in another memory $\cB$ if and only if $C(\cA) \subseteq C(\cB)$. 
In other words, $\cA$ bulk-embeds in $\cB$ if and only if it has no state
$\rho$ with too much entropy to fit in $\cB$.

Our second main result is the following noiseless coding theorem, which
generalizes a result of Barnum, Hayden, Jozsa, and Winter
\cite{BHJW:reversible}. The terms of the theorem and a self-contained proof
appear in \sec{s:coding}.

\begin{theorem} Let $\cA$ be a quantum memory with a state $\rho$ and let $\cB$
be another quantum memory.  Then there is a reliable noiseless
coding sequence
$$\begin{array}{c@{\hspace{2cm}}c@{\hspace{2cm}}c} \\[-1ex]
\rnode{a}{\cA^{\tensor N}} & \rnode{b}{\cB^{\tensor N(1+\eps)}}
& \rnode{c}{\cA^{\tensor N}}
\end{array}
\ncline[nodesep=.3]{->}{a}{b}\Aput{\cY_N}
\ncline[nodesep=.3]{->}{b}{c}\Aput{\cX_N}
$$
for every rational $\eps > 0$ if and only if $(H_\cA(\rho),S_\cA(\rho)) \in
C(\cB)$.  Here ``reliable'' means that the 
complete fidelity $F(\rho^{\tensor N},\cX_n \circ \cY_n) \to 1$
as $N \to \infty$.
\label{th:code} \end{theorem}

The ``no-go'' direction of \thm{th:code} depends on an interesting H\"older
inequality for fidelity of encodings, \thm{th:squeeze}.  In simplified
form, our inequality says that if
$$\begin{array}{c@{\hspace{1.5cm}}c@{\hspace{1.5cm}}c} \\[-1ex]
\rnode{a}{\cA} & \rnode{b}{\cB} & \rnode{c}{\cA}
\end{array}
\ncline[nodesep=.3]{->}{a}{b}\Aput{\cY}
\ncline[nodesep=.3]{->}{b}{c}\Aput{\cX}
$$
are two quantum operations and $\frac1p + \frac1q = 1$, then
$$\Tr(\cX \circ \cY) \le ||\lambda(\cA)||_q\;||\lambda(\cB)||_p.$$
This inequality is a broad generalization of the following elementary
combinatorial fact:  If a (uniformly) random number $x$ from 1 to $a$ is
encoded into a random number from 1 to $b$ with $b < a$ and decoded back again,
then the probability that $x$ is recovered is at most $\frac{b}{a}$.

In conclusion, \thm{th:code} is an important converse to \thm{th:embed}. 
Together they say that if $\cA$ and $\cB$ are two hybrid quantum memories,
then, then either $\cA$ blindly bulk-encodes into $\cB$ with perfect fidelity,
or $\cA$ has a state $\rho$ that does not visibly bulk-encode into $\cB$ with
high fidelity.

\section{Memory}
\label{s:memory}

As explained in the introduction, the first question is whether our
model of a hybrid memory is adequately general. One justification comes from
viewing a quantum system not as a Hilbert space, but as an abstract operator
algebra $\cA$.  If $\cA$ is infinite-dimensional, it should satisfy some
analytic axioms in order to be useful for quantum probability theory; usually
it is assumed to be either a $C^*$-algebra or a von Neumann algebra
\cite{KR:vol1,KR:vol2}.  But if it is finite-dimensional, it suffices to
require that $\cA$ be a (positive-definite) \emph{$*$-algebra}; it is then also
a $C^*$-algebra and a von Neumann algebra.  This means that in addition to the
fact that $\cA$ is a complex vector space with associative multiplication, it
has an abstract $*$-operation which is anti-linear, product-reversing, and
suitably positive-definite:
$$(\lambda A B)^* = \overline{\lambda}B^*A^*\qquad A^*A = 0 \implies A = 0.$$
Positive definiteness leads to an important partial ordering on
$\cA$.  By definition $X \ge Y$ if $X - Y = A^*A$ for some $A$.

For example, the matrix algebra $\cM_n$ is a $*$-algebra.

Despite their abstraction, $*$-algebras have all of the necessary structure for
quantum information theory.  The elements of a $*$-algebra $\cA$ of the form
$A^*A$ are called \emph{positive}.  A \emph{state} $\rho$ on a $*$-algebra
$\cA$ is defined as a dual vector $\rho \in \cA^*$ which is positive on
positive elements and which is normalized by $\rho(I) = 1$.  Consequently we
write $\rho(A)$ for the expectation of $A$ rather than $\Tr(\rho A)$. (The
latter notation is of course equivalent when $\cA$ is a matrix algebra; it
expresses $\rho$ as a \emph{density operator}.) A \emph{quantum operation}
from a system with $*$-algebra $\cA$ to a system with $*$-algebra $\cB$ is
defined as a unital, completely positive (UCP) linear map $\cE:\cA \to \cB.$
Here \emph{completely positive} means that $\cE$ sends positive elements to
positive elements after tensoring with the identity on a third $*$-algebra. 
Note that the transpose $\cE^T:\cB^* \to \cA^*$ is the corresponding map on
states.  It is completely positive and trace-preserving if we take $\rho(I)$ to
be the trace of $\rho$.

It will be useful to consider a larger class of maps than traditional quantum
operations.  A completely positive map $\cE:\cA \to \cB$ is \emph{subunital}
(or SUCP) if $\cE(I) \le I$. Whereas a UCP map conserves probability, an SUCP
map either conserves or diminishes it.   An SUCP map can be physically realized
in the same way as a UCP map, with the extra interpretation that missing
probability corresponds to ending the experiment.  An SUCP map can also be
called a \emph{decay quantum operation}.

A standard classification theorem \cite{Bratteli:inductive} says that every
finite-dimensional $*$-algebra $\cA$ is a direct sum of matrix algebras,
$$\cA \cong \bigoplus_{k=1}^n \cM_{\lambda_k}.$$
Thus a quantum memory of shape $\lambda$ is the most general possible
finite-dimensional complex algebra of observables satisfying reasonable
algebraic axioms.  (However abandoning $\C$ as the field of scalars leads to
other possibilities \cite{CFR:rebits}.)

Another justification comes from the interaction of a physical memory with its
environment. Consider a physical device whose state is defined by a $*$-algebra
$\cM$.  Realistically $\cM$ is very large, but almost all of it is thermally
coupled to the environment.  Its decoherence on the thermal time scale is given
by some decay quantum operation $\cE:\cM \to \cM$.  If the thermal time scale
is much shorter than the computational time scale, then the information
retained by $\cE^n$ in the limit $n \to \infty$ is the reliable memory of
$\cM$.

Certainly any finite-dimensional $*$-algebra $\cA$ is the reliable memory
retained by some quantum operation on a matrix algebra $\cM_d$.  In the minimal
construction, let $d = ||\lambda(\cA)||_1$ be the total size of all blocks of
$\cA$.  We realize $\cA \subseteq \cM_d$ as matrices with a diagonal block of
size $\lambda_k(\cA)$ for each $k$. The algebra $\cM_d$ has a POVM whose $k$th
element $P_k$ is the identity of the $k$th summand $\cA_k$.  The corresponding
quantum operation
$$\cP(A) = \sum_{k=1}^n P_k A P_k$$
is a projection, meaning $\cP^2 = \cP$, and its image is $\cA$. If the thermal
evolution of $\cM_d$ is given by $\cP$, the algebra $\cA$ measures the retained
information.

Conversely, the following two results show that if $\cE$ is a (decay) quantum
operation on a finite-dimensional $*$-algebra, the information retained by
$\cE^n$ in the limit $n \to \infty$ is measured by a smaller $*$-algebra of
effective observables. (See also Zurek \cite{Zurek:rules}.)

\begin{theorem} Let $\cE:\cM \to \cM$ be an SUCP map on a finite-dimensional
$*$-algebra $\cM$. Then there exists a sequence of integers $n_k \to \infty$
such that $\cE^{n_k}$ converges to a unique projection $\cP$.
\label{th:proj}
\end{theorem}

\begin{proof}(Sketch) Choose a basis of $\cM$ that puts $\cE$ in Jordan
canonical form. Since $\cE^n$ is SUCP, its matrix entries are bounded. 
Therefore $\cE$ has no eigenvalues $\lambda$ with $|\lambda| > 1$, and if
$|\lambda| = 1$, the $\lambda$-isotypic part of $\cE$ is diagonal.  Choose a
sequence of exponents $n_k \to \infty$ such that the phases of these diagonal
entries of $\cE^{n_k}$ are aligned with $1$ in the limit.  The rest of the
matrix of $\cE^n$ decays to $0$ as $n \to \infty$.  The map $\cP$ is unique
because if the phases do not align with 1, the limiting map is not a
projection.
\end{proof}

Finally a result of Choi and Effros \cite[pp.166-7]{CE:injectivity}
completes our justification for the $*$-algebra model.

\begin{theorem}[Choi, Effros] If $\cM$ is a finite-dimensional $*$-algebra and
$\cP$ is an SUCP projection on $\cM$, then the image of $\cP$ is a $*$-algebra
$\cA$ with a modified product $A \circ B = \cP(AB)$.
\label{th:ce1} \end{theorem}

The non-trivial part of \thm{th:ce1} (which more generally holds for
$C^*$-algebras) is the fact that the modified product $A \circ B$ is
associative.  The modified product structure is consistent with applying $\cP$
between any two computational manipulations of $\cM$.  Technically speaking,
Choi and Effros prove \thm{th:ce1} for UCP maps, but the proof for SUCP maps is
the same.

A quantum operation $\cX:\cB \to \cA$ is a \emph{blind, perfect-fidelity
encoding} if it has a right inverse $\cY:\cA \to \cB$, which is then called the
\emph{decoding}. In this case the reverse composition $\cY \circ \cX$ is a CPU
projection $\cP$.  Moreover, $\cY$ identifies $\cA$ with the Choi-Effros algebra
structure on $\cP$.  This construction is reversible: Given $\cP$, we can
define $\cA$ to be $\im \cP$ with its Choi-Effros structure. Certainly if $\cY$
embeds $\cA$ into $\cB$, then a corresponding $\cX$ exists.  (If $\cY$ is not
unital, then it is a decay quantum operation, but $\cX$ can always be made
non-decay.) Generally, even when $\cA$ and $\cB$ are abelian, $\cY$ is not an
algebra embedding, but another argument of Choi and Effros
\cite[pp.202-3]{CE:injectivity} says that it always yields one.

\begin{theorem}[Choi, Effros] If $\cM$ is a finite-dimensional
$*$-algebra and $\cP$ is an SUCP projection on $\cM$, then 
$\im cP$ also embeds (non-unitally) as a subalgebra of $\cM$.
\label{th:ce2} \end{theorem}

\thm{th:ce2} more generally holds for von Neumann algebras. The proof adjusts
$\cP$ in a canonical way.  It is not hard to show that every algebra embedding
is a blind, perfect-fidelity decoding $\cY$; there exists an $\cX$ to match it.

\section{Embeddings}
\label{s:embed}

\subsection{Bin packing}
\label{s:bin}

Besides embeddability and bulk embeddability, we will also compare memories
using a partial ordering on partitions which resembles dominance
\cite[Ch.7]{Stanley:enumerative2}, or majorization, but is stricter. The
partition $\lambda$ \emph{supermajorizes} the partition $\mu$, or $\mu \pe_S
\lambda$, if for every $n$, the sum of all parts of $\lambda$ that are
at least $n$ exceeds the same sum for $\mu$.  \lem{l:fudge} below and
\thm{th:embed} imply that supermajorization lies between embeddability and bulk
embeddability:
\begin{gather*}
\cA \injects \cB \implies \lambda(\cA) \pe_S \lambda(\cB)
\implies \cA \binjects \cB \\
\cA \binjects \cB \nimplies \lambda(\cA) \pe_S \lambda(\cB)
\nimplies \cA \injects \cB.
\end{gather*}

We can view the parts of a partition $\lambda$ as an unordered multiset
$\{\lambda_k\}$.  It is sometimes convenient to assume a specific order on
the parts.  In this case we follow the usual convention that the parts of
$\lambda$ are non-increasing:
$$\lambda_1 \ge \lambda_2 \ge \cdots \ge \lambda_n \ge 1.$$

Given a partition $\lambda$, let $\lambda_{\ge x}$ denote the sum of all parts
of $\lambda$ that are at least $x$. Thus $\lambda \pe_S \mu$ means that
$$\lambda_{\ge x} \le \mu_{\ge x}$$
for all $x$.  Obviously integer values of $x$ suffices, but
it will be convenient later to allow non-integer values.
Also $\ell\lambda$ denotes $\lambda$ with each part repeated $\ell$ times.
(This is not to be confused with magnifying each part by a factor of $\ell$.)

In order to analyze bulk embeddings and prove \thm{th:embed}, we first analyze
ordinary embeddings \cite{Bratteli:inductive}.  If $\cA$ and $\cB$ are
finite-dimensional $*$-algebras, then any algebra homomorphism $f:\cA \to \cB$
is characterized by a \emph{Bratteli diagram} $\Gamma$ whose vertices are the
summands of $\cA$ and $\cB$.  Let $\cA_k$ be the $k$th summand of $\cA$, so
that $\cA_k \cong \cM_{\lambda_k}$, and likewise for $\cB$. If we denote the
adjacency matrix of $\Gamma$ by $\Gamma$ as well, then the diagram's
interpretation is that $f$ embeds $\Gamma_{j,k}$ copies of $\cA_j$ in $\cB_k$.
(The matrix $\Gamma$ is the adjacency matrix of the diagram $\Gamma$.)  The
matrix $\Gamma$ must satisfy the inequality
$$\sum_j \Gamma_{j,k}\lambda(\cA)_j \le \lambda(\cB)_k$$
for all $k$.  (Bratteli diagrams often describe unital
homomorphisms, which require equality.) The homomorphism $f$ is an embedding if
and only if each summand of $\cA$ has at least one edge, or equivalently that
$$\sum_k \Gamma_{j,k} \ge 1$$
for all $j$.

Thus we can think of $\cA$ as a set of 1-dimensional blocks, $\cB$ as a set of
1-dimensional bins, and the embedding as a way to pack the blocks of $\cA$ in
the bins of $\cB$.  The packing might repeat some of the summands of $\cA$, but
if there is any embedding, there is one with no repetition.  (Repetition in
this sense has nothing to do with cloning as in the no-cloning theorem.  In
representation theory this kind of repetition is usually called
\emph{multiplicity}.)

\begin{lemma} If $\cA \injects \cB$, then
$\lambda(\cA) \pe_S \lambda(\cB)$. If
$2\lambda(\cA) \pe_S \lambda(\cB)$, then $\cA \injects \cB$.
\label{l:fudge} \end{lemma}

\begin{proof} Both statements follow by induction on the number of parts of
$\lambda(\cA)$.  They both hold trivially when $\lambda(\cA)$ is empty.  To
prove the first assertion, suppose that in some embedding, $\cA_1$ embeds in
$\cB_k$.  Let $\hcA$ be $\cA$ with $\cA_1$ removed and let $\hcB$ be $\cB$ with
$\cB_k$ reduced by $\lambda(\cA)_1$, or removed if $\lambda(\cB)_k =
\lambda(\cA)_1$.  By construction, $\hcA \injects \hcB$.  Thus by
induction,
$$\lambda(\hcA)_{\ge x} \le \lambda(\hcB)_{\ge x}$$
for all $x \ge 1$.  By the definition of $\hcA$ and $\hcB$,
\begin{align*}
\lambda(\hcA)_{\ge x} &= \lambda(\cA)_{\ge x} - \lambda(\cA)_1 \\
\lambda(\hcB)_{\ge x} &\le \lambda(\cB)_{\ge x} - \lambda(\cA)_1
\end{align*}
for $x \le \lambda(\cA)_1$, while $\lambda(\cA)_{\ge x}$ vanishes for $x >
\lambda(\cA)_1$.  Thus
$$\lambda(\cA)_{\ge x} \le \lambda(\cB)_{\ge x},$$
as desired.

To prove the second assertion, suppose that $2\lambda(\cA) \pe_S \lambda(\cB)$,
or equivalently that
$$2\lambda(\cA)_{\ge x} \le \lambda(\cB)_{\ge x}$$
for all $x$.  We can greedily put $\cA_1$ in any $\cB_k$ in which it fits and
make $\hcA$ and $\hcB$ as before.  (In this greedy algorithm it is
important to start with the largest summand of $\cA$, not an arbitrary one.) If
$\lambda(\cB)_k \le 2\lambda(\cA)_1$, then
\begin{align*}
\lambda(\hcA)_{\ge x} &= \lambda(\cA)_{\ge x} - \lambda(\cA)_1 \\
\lambda(\hcB)_{\ge x} &\ge \lambda(\cB)_{\ge x} - 2\lambda(\cA)_1
\end{align*}
for all $x \le \lambda(\cA)_1$, while $\lambda(\hcA)_{\ge x}$ vanishes for
$x > \lambda(\cA)_1$.  On the other hand if $\lambda(\cB)_k \ge
2\lambda(\cA)_1$, then bin $k$ remains larger than any block even after block 1
is subtracted. In this case
\begin{align*}
\lambda(\hcA)_{\ge x} &= \lambda(\cA)_{\ge x} - \lambda(\cA)_1 \\
\lambda(\hcB)_{\ge x} &\ge \lambda(\cB)_{\ge x} - \lambda(\cA)_1
\end{align*}
for all $x \le \lambda(\cA)_1$. Thus
$$2\lambda(\hcA) \pe_S \lambda(\hcB)$$
either way, so the bin packing exists by induction.
\end{proof}

\subsection{Large deviations}
\label{s:large}

The proof of \thm{th:embed} combines \lem{l:fudge} with the Chernoff-Cram\'er
theorem on large deviations \cite{DZ:techniques}.  The theorem is usually
stated in terms of sums of independent random variables, but it is more
convenient here to formulate it in terms of convolutions of measures.

\begin{theorem}[Chernoff, Cram\'er] Let $\mu$ be a measure on an
interval $[0,u]$, let
$$\ell(\beta) = \log \int_0^\infty e^{\beta x} d\mu(x)$$
be the logarithm of the Laplace transform of $\mu$ and let $t>0$. Then for all
$n \in \Z_+$ and all $\beta > 0$,
$$\int_{nt}^{\infty} d\mu^{*n} \le e^{n(\ell(\beta) - \beta t)}$$
If $\ell'(0) \le t < u$ and $\beta$ minimizes
$$\ell(\beta) - \beta t,$$
then for all $0 < s < t$,
$$\int_{n(t-s)}^{\infty} d\mu^{*n} \ge e^{n(\ell(\beta) - \beta t - \beta s)}
    \biggl( 1 - \frac{\ell''(\beta)}{ns^2}\biggr).$$
\label{th:cc} \end{theorem}

Here $\mu^{*n}$ denotes the $n$-fold convolution of $\mu$ with itself.
When $\ell'(0) < t < u$, the expression
$$\hat{\ell}(t) = \min_{\beta} \ell(\beta) - \beta t$$
is the Legendre transform of $\ell(\beta)$.  Note that a unique $\beta$
achieves the minimum because the minimand is concave up, increases as $\beta
\to \infty$, and does not increase at $\beta = 0$.

\begin{proof}(Sketch) For any $\beta$,
\begin{align*}
\int_{nt}^{\infty} d\mu^{*n}
    &\le e^{-n\beta t} \int_0^\infty e^{\beta x} d\mu^{*n}(x) \\
    &= e^{- n\beta t} e^{n\ell(\beta)}.
\end{align*}
This establishes the upper bound, Chernoff's inequality.

If $\beta$ is chosen to minimize $\ell(\beta) - \beta t$, then $t =
\ell'(\beta)$.  In this case
\begin{align*}
\int_{n(t-s)}^{\infty} d\mu^{*n}
    &\ge e^{-n\beta(s+t)} \int_{n(t-s)}^{n(t+s)} e^{\beta x} d\mu^{*n}(x) \\
    &\ge e^{-n\beta(s+t)} \int_0^\infty
        \biggl(1 - \frac{(x-nt)^2}{(ns)^2}\biggr) e^{\beta x} d\mu^{*n}(x) \\
    &= e^{-n\beta(s+t)} \biggl(1 - \frac{\ell''(\beta)}{ns^2}\biggr)
        e^{n\ell(\beta)}.
\end{align*}
The equality uses the identities
\begin{align*}
\int_0^\infty x e^{\beta x} d\mu^{*n}(x)  &= \bigl(e^{n\ell(\beta)}\bigr)'
    = n\ell'(\beta) e^{n\ell(\beta)} \\
\int_0^\infty x^2 e^{\beta x} d\mu^{*n}(x) &= \bigl(e^{n\ell(\beta)}\bigr)''
    = \bigl(n\ell''(\beta) + n^2\ell'(\beta)^2\bigr) e^{n\ell(\beta)}.
\end{align*}
This establishes the lower bound, Cram\'er's theorem.
\end{proof}

\begin{proof}[Proof of \thm{th:embed}] In brief, without loss of
generality
$$||\lambda(\cA)||_p < ||\lambda(\cB)||_p$$
for all $p \in [1, \infty]$.  In this case we apply
\thm{th:cc} to the measures
\begin{align*}
\mu_\cA &= \sum_k \lambda_k(\cA) \delta_{\log \lambda_k(\cA)} \\
\mu_\cB &= \sum_k \lambda_k(\cB) \delta_{\log \lambda_k(\cB)},
\end{align*}
where $\delta_x$ denotes a delta function (or atom) at $x$. For sufficiently
large $n$, Chernoff's bound for $\mu_\cA$ and Cram\'er's inequality for
$\mu_\cB$ together imply the criterion
$$2\lambda(\cA^{\tensor n})_{\ge x} \le \lambda(\cB^{\tensor n})_{\ge x}$$
of \lem{l:fudge} uniformly for $x \in [1,\infty)$.

In detail, we assume that $||\lambda(\cB)||_\infty > 1$;
otherwise $\cA$ and $\cB$ are both entirely classical and \thm{th:embed}
is easy.  Since
$$||\lambda(\cA)||_p \le ||\lambda(\cB)||_p$$
for all $p \in [1, \infty]$, then for any $k > 1$,
$$||\lambda(\cA^{\tensor k})||_p < ||\lambda(\cB^{\tensor k+1})||_p.$$
The $\eps$ margin in \thm{th:embed} thus allows us to assume that
$$||\lambda(\cA)||_p < ||\lambda(\cB)||_p$$
for all $p \in [1, \infty]$ by replacing $\cA$ by
$\cA^{\tensor k}$ and $\cB$ by $\cB^{\tensor k+1}$.

The measure
$\mu_\cA$ is defined so that
$$\mu_\cA^{*n} = \mu_{\cA^{\tensor n}}$$
and
$$\lambda(\cA)_{\ge e^x} = \int_x^\infty d\mu_\cA(x),$$
and likewise for $\mu_\cB$.  Therefore by \lem{l:fudge}, it
suffices to show that there exists an $n$ such that
for all $t \ge 0$,
\eq{e:cutoff}{2\int_{nt}^\infty d\mu_\cA^{*n}
    \le \int_{nt}^\infty d\mu_\cB^{*n}.}

As in the statement of \thm{th:cc}, let
\begin{align*}
\ell_\cA(\beta) = \log \int_0^\infty e^{\beta x} d\mu_\cA(x)
    = \log ||\lambda(\cA)||_{\beta+1}^{\beta+1} \\
\ell_\cB(\beta) = \log \int_0^\infty e^{\beta x} d\mu_\cB(x)
    = \log ||\lambda(\cB)||_{\beta+1}^{\beta+1}.
\end{align*}
Observe that $\ell_\cB(\beta)$ is a smooth, concave function, and that
$$\lim_{\beta \to \infty} \frac{\ell'_\cB(\beta)}{\beta}
    = \log ||\lambda(\cB)||_\infty < \infty.$$
It follows that $\ell''_\cB(\beta)$ has a finite maximum $C$ for $\beta \in
[0,\infty)$.
Note also that
$$\frac{\ell_\cB(\beta) - \ell_\cA(\beta)}{\beta}$$
achieves a positive minimum, since
\begin{align*}
\lim_{\beta \to \infty} \frac{\ell_\cB(\beta) - \ell_\cA(\beta)}{\beta}
    &= ||\lambda(\cB)||_\infty - ||\lambda(\cA)||_\infty \\
\lim_{\beta \to 0} \frac{\ell_\cB(\beta) - \ell_\cA(\beta)}{\beta} &= \infty.
\end{align*}

Temporarily suppose that $t \ge \ell'_\cB(0)$ and that $\beta = \beta(t)$
minimizes $\ell_\cB(\beta) - \beta t$.  Let
$$s = \sqrt{\frac{2C}{n}}.$$
Then
\begin{align*}
\int_{n(t-s)}^{\infty} d\mu_\cA^{*n} &\le
    e^{n(\ell_\cA(\beta) - \beta t + \beta s)} \\
\int_{n(t-s)}^{\infty} d\mu_\cB^{*n} &\ge
    e^{n(\ell_\cB(\beta) - \beta t - \beta s) - \log 2}.
\end{align*}
If $n$ is large enough that
$$2s + \frac{2\log 2}n = 2\sqrt{\frac{2C}{n}}+ \frac{2\log 2}n
    \le \min_{\beta} \frac{\ell_\cB(\beta) - \ell_\cA(\beta)}{\beta},$$
then
$$2\int_{n(t-s)}^{\infty} d\mu_\cA^{*n}
    \le \int_{n(t-s)}^{\infty} d\mu_\cB^{*n}.$$
Thus for some $\eps > 0$, inequality \eqref{e:cutoff} holds
for all $t > \ell'_\cB(0) - \eps$.

If $t \le \ell'_\cB(0) - \eps$, let $u = \ell'_\cB(0)$ and let $\beta = 0$.
Then
$$\int_{nt}^{\infty} d\mu_\cA^{*n} \le \int_0^{\infty}
    d\mu_\cA^{*n} = e^{n\ell_\cA(0)},$$
while
$$\int_{nt}^{\infty} d\mu_\cB^{*n} \ge
\int_{n(u-s)}^{\infty} d\mu_\cB^{*n} \ge
    e^{n\ell_\cB(0) - \log 2}$$
provided that $s \le \eps$.  Since $\ell_\cA(0) < \ell_\cB(0)$, inequality
\eqref{e:cutoff} holds when $n$ is large enough.
\end{proof}

\section{Entropy}
\label{s:entropy}

\subsection{Capacity}
\label{s:capacity}

Let $\cA$ be a finite-dimensional $*$-algebra, where as before
$$\cA = \bigoplus_{k=1}^n \cA_k \cong \bigoplus_{k=1}^n \cM_{\lambda_k}.$$
Let $\rho$ be a (mixed) state on $\cA$; as explained above we view
$\rho$ as a dual vector on $\cA$ rather than as an element of $\cA$.  Let 
$$\rho_k = \rho|_{\cA_k}$$
be the restriction of $\rho$ to $\cA_k$.  Diagonalize each $\rho_k$ and let
$r_{k,j}$ with $1 \le j \le \lambda_k$ be its diagonal entries. (In general a
state $\rho$ on matrices is diagonal if and only if $\rho(A)$ depends only on
the diagonal entries of $A$.  Equivalently in the present case we can interpret
$\rho$ as a density operator.) Let
$$r_k = \rho_k(I) = \sum_{j=1}^{\lambda_k} r_{k,j}$$
be the total density of $\rho$ in $\cA_k$; evidently
$$\sum_{k=1}^n r_k = 1.$$
We also define the normalized state $\rho'_k$ on
$\cA_k$ by 
$$\rho'_k = \frac{\rho_k}{r_k},$$
with diagonal entries
$$r'_{k,j} = \frac{r_{k,j}}{r_k}.$$

The \emph{classical entropy} of the state $\rho$ on $\cA$ is defined as
$$H_\cA(\rho) = -\sum_{k=1}^n r_k \log\ r_k.$$
The \emph{quantum entropy} of $\rho$ is defined as 
$$S_\cA(\rho) = -\sum_{k=1}^n \sum_{j=1}^{\lambda_k} r_{k,j} \log\ r'_{k,j}.$$
(Note that in the literature $H$ is also sometimes used to denote
quantum, or von Neumann, entropy.  Here we follow the convention of
Nielsen and Chuang \cite{NC:book}.)
These two entropies are supported by a number of elementary justifications: The
classical entropy of $\rho$ is the Shannon entropy of the restriction of $\rho$
to the center of $\cA$, which is a classical system.  The quantum entropy of
$\rho$ is the expected value of the von Neumann entropy of $\rho_k$, where the
index $k$ is chosen randomly with probability $r_k$. Finally the total entropy
$$H_\cA(\rho) + S_\cA(\rho) =
    -\sum_{k=1}^n \sum_{j=1}^{\lambda_k}r_{k,j} \log\ r_{k,j}$$
has the same formula as both the Shannon and the von Neumann entropy.

The proof of \thm{th:cap} is based on finding thermal states of $\cA$ with
respect to a certain Hamiltonian.  We define the energy $E_k$ of the summand
$\cA_k$ as the negative of its capacity for quantum entropy:
$$E_k = - \log \lambda_k(\cA).$$
We retain the parameter $\beta$ from \sec{s:large}, setting $p = \beta+1$, and
we also define the temperature $T = 1/\beta$. The thermal state $\rho_T$ at
temperature $T$ has the property that its restriction $\rho_k$ to each $\cA_k$
is uniform.  If $\rho$ is any state with this property, then its energy
$E_\cA(\rho)$ is, by definition, the negative of its quantum entropy:
$$E_\cA(\rho) = -S_\cA(\rho).$$
The free energy of $\rho$ is therefore
\begin{align*}
F_\cA(\rho) &= E_\cA(\rho) - T(H_\cA(\rho)+S_\cA(\rho)) \\
    &= -T(H_\cA(\rho)+pS_\cA(\rho)).
\end{align*}
Since the thermal state minimizes the free energy, we have defined energy so
that the thermal state $\rho_T$ maximizes quantum entropy plus classical
entropy discounted by $p$. To compute the maximum, recall that for the thermal
state $\rho_T$, the free energy is proportional to the log of the partition
function:
\begin{align*}
F_\cA(\rho_T) &= -T\log\ Z_\cA(\rho_T) = -T\log\
        \biggl(\sum_{k=1}^n \lambda_k e^{\beta\log\ \lambda_k(\cA)}\biggr) \\
    &= -T\log \sum_{k=1}^n \lambda_k^{\beta+1} = -Tp\log\ ||\lambda(\cA)||_p.
\end{align*}
Therefore
$$\frac{H_\cA(\rho_T)}{p}+S_\cA(\rho_T) = \log\ ||\lambda(\cA)||_p,$$
as desired.

To prove the final claim of \thm{th:cap}, observe that every point in
$C(\cA)$ can be written in the form
$$(H_\cA(\rho_T)+t,S_\cA(\rho_T) - s - t)$$
with $0 \le s,t$ and $s+t \le S_\cA(\rho_T)$.  Starting with the
state $\rho_T$, the quantum entropy in each block
can be decreased to $0$ without changing the total
probability of that block, hence without changing the classical
entropy.  In this way we can absorb the constant $s$.
The remaining constant $t$ just matches the one in the conclusion.

\subsection{Noiseless coding}
\label{s:coding}

A final justification for quantum and classical entropies is \thm{th:code},
which we prove here. The theorem is a mutual generalization of, and entirely
analogous to, Shannon's classical and Schumacher's purely quantum coding
theorems \cite[Thms. 12.4 \& 12.6]{NC:book}
\cite{Shannon:theory,Schumacher:coding}.

Given an algebra $\cA$ with a state $\rho$ and a second algebra $\cB$,  a
\emph{noiseless coding} is a pair of decay quantum operations
$$\begin{array}{c@{\hspace{1.5cm}}c@{\hspace{1.5cm}}c} \\[-1ex]
\rnode{a}{\cA} & \rnode{b}{\cB} & \rnode{c}{\cA}
\end{array}
\ncline[nodesep=.3]{->}{a}{b}\Aput{\cY}
\ncline[nodesep=.3]{->}{b}{c}\Aput{\cX}.
$$
Since these are maps on algebras rather than states spaces, the second map
$\cX$ is the encoding and the first map $\cY$ is the decoding.

We are interested in reliable noiseless coding, or in other words
high-fidelity, visible bulk-encoding. But a rigorous definition of reliability
is not obvious. Suppose that $\cE$ is a decay quantum operation from a memory
$\cA$ to itself, and that $\cA$ has a state $\rho$.  If $\cC$ is another
memory, we define the \emph{$\cC$-fidelity} of $\cA$ to be
\eq{e:fidel}{F_{\cC}(\rho,\cE) =
    \min_{\substack{\sigma \in (\cC \tensor \cA)^* \\ \sigma \mapsto \rho}}
    1-D\bigl(\sigma,(\mathrm{id.} \tensor \cX^T)(\sigma)\bigr),}
where $D$ is the trace distance on states, and the minimum is taken over states
$\sigma$ on $\cC \tensor \cA$ that project to the state $\rho$ on $\cA$.  In
words, the $\cC$-fidelity is the complement of the highest probability that the
operation $\cX$ leaves the larger system $\cC \tensor \cA$ in an erroneous
state. We define the \emph{complete fidelity} $F(\rho,\cE)$ to be the infimum
of $\cC$-fidelity over all $\cC$.  It is not hard to show that complete
fidelity agrees with the classical non-error rate when $\cA$ is classical, and
with entanglement fidelity when $\cA$ is purely quantum.

The more difficult half of \thm{th:code} is the no-go direction. To review, the
heart of the no-go direction of the classical encoding theorem is the following
elementary fact about squeezing states:  If a state $\rho$ of a classical
memory is encoded into $b$ values, then it cannot be recovered with probability
greater than $b||\rho||_\infty$, where $||\rho||_\infty$ is the probability of
the most likely value of $\rho$.  Or for simplicity, if $\rho$ is the uniform
state on a memory with $a$ values, then the non-error rate is at most
$\frac{b}{a}$. We will need a hybrid quantum generalization of this
inequality.  To state it, we replace $||\rho||_\infty$ with a different norm.
If $\rho$ is a state on $\cA$, define the \emph{dense-coding-based
supremum} of $\rho$ by
$$||\rho||_\d = \max_{j,k} \frac{r_{k,j}^2}{r_k}$$
in the notation of \sec{s:capacity}.  

\begin{theorem} Let $\cA$ and $\cB$ be two hybrid quantum memories and let
$\rho$ be a state on $\cA$. If
$$\begin{array}{c@{\hspace{1.5cm}}c@{\hspace{1.5cm}}c} \\[-1ex]
\rnode{a}{\cA} & \rnode{b}{\cB} & \rnode{c}{\cA}
\end{array}
\ncline[nodesep=.3]{->}{a}{b}\Aput{\cY}
\ncline[nodesep=.3]{->}{b}{c}\Aput{\cX}
$$
are decay quantum operations and $\frac1p + \frac1q = 1$, then
$$F(\rho,\cX \circ \cY) \le
    ||\rho||_\d\;||\lambda(\cA)||_q\;||\lambda(\cB)||_p.$$
\label{th:squeeze} \eatline \end{theorem}

Before proving \thm{th:squeeze}, we discuss some special cases. If $\cA = \C^a$
is classical and $\rho$ is the uniform state, then $||\rho||_\d = \frac1a$.  In
this case, taking $p=1$, \thm{th:squeeze} says that
$$F(\rho,\cX \circ \cY) \le \frac{||\lambda(\cB)||_1}{a}.$$
This generalizes the classical squeezing result, bounding the fidelity by the
total number of independent states of $\cB$ whether or not it is classical.  On
the other hand, if $\cA = \cM_a$ is purely quantum and $\rho$ is the uniform
state, then $||\rho||_\d = \frac1{a^2}$.  In this case, taking 
$p=\infty$, \thm{th:squeeze} says that
$$F(\rho,\cX \circ \cY) \le \frac{\lambda_1(\cB)}{a}.$$
In other words, if $\cA$ is purely quantum, then the fidelity of squeezing is
bounded by the largest quantum block of $\cB$, regardless of its classical
capacity.  (But if $\cB = \cM_b$ is also purely quantum, then it can be shown
that
$$F(\rho,\cX \circ \cY) \le \frac{b^2}{a^2}$$
when $\rho$ is inform. In this case if $b$ divides $a$, then multiplying $\cB$
by $\frac{a}{b}$ classical states can boost fidelity to $\frac{b}{a}$.)

\begin{proof} The operations $\cX$ and $\cY$ admit Kraus
representations
\begin{align*}
\cX(\bigoplus_k B_k) &=
    \bigoplus_j \sum_{k,\ell} X_{j,k,\ell}^* B_k X_{j,k,\ell} \\
\cY(\bigoplus_k A_k) &=
    \bigoplus_j \sum_{k,\ell} Y_{j,k,\ell}^* A_k Y_{j,k,\ell}
\end{align*}
subject to the subunital conditions
\begin{align}
\sum_{k,\ell} X_{j,k,\ell}^* X_{j,k,\ell} &\le I \in \cA_j
    \nonumber \\
\sum_{k,\ell} Y_{j,k,\ell}^* Y_{j,k,\ell} &\le I \in \cB_j.
    \label{e:u}
\end{align}
Recall the definition of $r_k$ and $\rho'_k$ in \sec{s:capacity}. The minimum
in equation \eqref{e:fidel} is obtained by lifting the state $\rho$ to the
completely correlated, completely entangled state
$$\sigma = \bigoplus_k r_k \psi_k$$
on $\cA \tensor \cA$, where $\psi_k$ is a pure state that projects to
$\rho'_k$. By a computation similar to one in Nielsen and Chuang \cite[p.
421]{NC:book}, the fidelity is then given by
\eq{e:fidelsum}{F = F(\rho,\cX \circ \cY) =
    \sum_{j,k,\ell,m} r_k|\rho'_k(Y_{j,k,m}X_{k,j,\ell})|^2.}

Given any state $\sigma$ on the matrix algebra $\cM_a$ and any matrices $X \in
\cM_{b \times a}$ and $Y \in \cM_{a \times b}$, the Cauchy-Schwarz inequality
and positivity together say that
$$|\sigma(YX)|^2 \le \sigma(X^*X)\sigma(YY^*) \le
    ||\sigma||_\infty^2\; \Tr(X^*X)\Tr(Y^*Y).$$
Applying this to equation \eqref{e:fidelsum}, we obtain the bound
\eq{e:bound}{F \le \sum_{j,k,\ell,m} ||\rho||_\d\;
    \Tr(X_{k,j,\ell}^* X_{k,j,\ell}) \Tr(Y_{j,k,m}^* Y_{j,k,m}).}
Define the numbers
$$x_{k,j} = \sum_\ell \Tr(X_{k,j,\ell}^* X_{k,j,\ell})
\qquad y_{j,k} = \sum_m \Tr(Y_{j,k,m}^* X_{j,k,m})$$
and define the vectors $x = (x_{k,j})$ and $y = (y_{j,k})$.
Then we can restate inequality \eqref{e:bound} as
$$F \le ||\rho||_\d\; \sum_{j,k} x_{k,j} y_{j,k} = ||\rho||_\d\; x \cdot y,$$
while equation \eqref{e:u} implies that
$$\sum_j x_{k,j} \le \lambda_k(\cA) \qquad
\sum_k y_{j,k} \le \lambda_j(\cB).$$
Thus
$$||x||_p \le ||\lambda(\cA)||_p \qquad ||y||_q \le ||\lambda(\cB)||_q.$$
Finally the H\"older inequality yields
\begin{align*}
F &\le ||\rho||_\d\; x \cdot y \le
    ||\rho||_\d\;||x||_p\;||y||_q \\
    &\le
    ||\rho||_\d\;||\lambda(\cA)||_p\;||\lambda(\cB)||_q
\end{align*}
when $\frac1p + \frac1q = 1$, as desired.
\end{proof}

\begin{proof}[Proof of \thm{th:code}](Semi-sketch) As in the proofs of
Shannon's and Schumacher's theorems as presented by Nielsen and Chuang
\cite{NC:book}, we first establish the existence of a $\eps$-typical subalgebra
$\cA_\typ$ of $\cA^{\tensor N}$ with respect to the state $\rho$.  (The $\eps$
in the proof here is not the same as the one in the statement of the theorem,
which we rename $\delta$.) We will take $\eps$ to implicitly depend on $N$ with
$\eps \to 0$ slowly as $N \to \infty$.  We will establish that $\cA_\typ$ is
approximately rectangular and that the restriction $\rho_\typ$ of
$\rho^{\tensor N}$.  We will then confirm that if
$$(S,H) = (S_\cA(\rho),H_\cA(\rho)) \in C(\cB),$$
then $\cA_N$ embeds in $\cB^{\tensor N}$ for sufficiently
large $N$; in particular it reliably encodes.  On the other hand,
if $(S,H) \not\in C(\cB)$, we will confirm that
$\cA_N$ does not reliably encode in $\cB^{\tensor N}$; indeed
the fidelity of any encoding-decoding converges to $0$ exponentially.

Assume that the state $\rho$ on $\cA$ is diagonalized and that $r_{k,j}$, with
$1 \le k \le n(\cA)$ and $1 \le j \le \lambda_k(\cA)$, are its diagonal
entries.  Here $n(\cA)$ denotes the number of parts of $\lambda(\cA)$.  This
induces a diagonalization of the state $\rho^{\tensor N}$ with a diagonal entry
$r_{K,J}$ for each pair of \emph{admissible} sequences
$$K = (k_1,k_2,\ldots,k_N) \qquad J = (j_1,j_2,\ldots,J_N)$$ is
such that
$$1 \le \ell \le N \qquad 1 \le k_\ell \le n(\cA)
    \qquad 1 \le j \le \lambda_{k_\ell}(\cA).$$
Moreover, for each admissible $K$, $\cA^{\tensor N}$ has an algebra summand
$(\cA^{\tensor N})_K$.  If $(K,J)$ and $(K,J')$ are two admissible pairs,
the algebra summand $(\cA^{\tensor N})_K$ has an elementary
matrix $E_{K,J,J'}$; these matrices then form a basis of
$\cA^{\tensor N}$.  We will consider a set $T$ of admissible pairs $(K,J)$
called the \emph{typical set}; momentarily it can be any set.
The span of the matrices $E_{K,J,J'}$ with $(K,J),(K,J') \in T$
is a subalgebra $\cA_\typ$.  Another way to describe the algebra $\cA_\typ$
is to define the projector
$$P_\typ = \sum_{(K,J) \in T} E_{K,J,J}$$
and then let
$$\cA_\typ = P_\typ \cA^{\tensor N} P_\typ.$$
In this notation, the map
$$\cP(X) = P_\typ X P_\typ$$
is an SUCP projection on $\cA^{\tensor N}$ with image $\cA_\typ$.

Given $\alpha > 0$, say that an admissible pair $(K,J)$ is \emph{$\alpha$-typical}
if the number of occurrences $N(K,J;k,j)$ of $(k,j)$ satisfies
$$\biggl|\frac{N(K,J;k,j) - r_{i,j}}N\biggr| < \alpha.$$
Let $T$ be the set of all $\alpha$-typical pairs.
By repeated application of Chernoff's inequality (\thm{th:cc} in 
a more traditional probabilistic context),
$$\rho^{\tensor N}(P_\typ) = \sum_{(K,J) \in T} r_{K,J} \to 1$$
for any fixed $\alpha$ as $N \to \infty$.  Moreover
$$F(\cP_\typ,\rho) \ge \rho^{\tensor N}(P_\typ)^2,$$
so for any fixed $\alpha$, $\cA_\typ$ and $\cA^{\tensor N}$ reliably encode
into each other.  At the same time, by a messy but straightforward calculation,
if $\alpha$ is sufficiently small relative to $\eps$ (and depending on $\rho$
but not on $N$), $\cA_\typ$ and $\rho_\typ$ have the following properties:
\begin{gather}
\bigl|(\log\ n(\cA_\typ)) - HN\bigr| < N\eps \nonumber \\
\bigl|(\log\ \lambda(\cA_\typ)_K) - HS\bigr| < N\eps \label{e:approx} \\
(\log\ ||\rho_d||) + H + 2S < N\eps. \nonumber
\end{gather}

Suppose that $(H,S) \in C(\cB)$.  In this case, let $C = e^{N(S+\eps)}$; then
$$\lambda(\cA_\typ)_{\ge C} = 0.$$
Meanwhile equations~\eqref{e:approx} imply that
$$\lambda(\cA_{N,\eps})_{\ge 0} < e^{N(H+S+2\eps)}.$$
By a derivation using Cr\'amer's bound like the one in the proof of
\thm{th:embed},
$$\lambda(\cB^{\tensor N(1+\delta)})_{\ge C} > 2e^{N(H+S+\eps)}$$
when $N$ is large enough, provided that $\eps$ is small compared to $\delta$.
Thus by \lem{l:fudge}, $\cA_{N,\eps}$ embeds in $\cB^{\tensor N(1+\delta)}$ for
large enough $N$, as desired.

Suppose that $(H,S) \not\in C(\cB)$.  In this case, suppose that
$$\begin{array}{c@{\hspace{1.5cm}}c@{\hspace{1.5cm}}c} \\[-1ex]
\Rnode{a}{\cA_\typ} & \Rnode{b}{\cB^{\tensor N(1+\delta)}}
    & \Rnode{c}{\cA_\typ}
\end{array}
\ncline[nodesep=.3]{->}{a}{b}\Aput{\cY}
\ncline[nodesep=.3]{->}{b}{c}\Aput{\cX}
$$
are decay quantum operations and that $\frac1p +
\frac1q = 1$.  By the first two equations of \eqref{e:approx},
$$\log\ ||\lambda(\cA_\typ)||_q < (\frac{H}q + S + 2\eps)N.$$
Combining this with \thm{th:squeeze} and the last equation of \eqref{e:approx},
we obtain
\thm{th:squeeze},
\begin{multline*}
\log\ F(\rho_\typ,\cX \circ \cY) < (\eps - H - 2S)N \\
 + (\frac{H}q + S + 2\eps)N
 + \log\ ||\lambda(\cB^{\tensor N(1+\delta)})||_p \\
= N((1+\delta)\log\ ||\lambda(\cB)||_p - \frac{H}p - S + 3\eps).
\end{multline*}
Since $\delta$ must be sent to $0$ and $\eps$ may be sent to $0$, the fidelity
therefore decays exponentially if there exists a $p$ such that
$$\log\ ||\lambda(\cB)||_p < \frac{H}p + S.$$
By the definition of $C(\cB)$, this inequality is equivalent to the assumed
condition $(H,S) \not\in C(\cB)$.   Since $F(\rho_\typ,\cX \circ \cY)$ decays
exponentially, it cannot converge to 1.
\end{proof}

\section{Discussion}

\sec{s:memory} illustrates the principle that classical information theory is
the abelian special case of quantum information theory.  Many authors maintain
a dichotomy between the two theories by considering ensembles of mixed states. 
But such formalism is ultimately redundant, because an ensemble is itself a
classical probabilistic state.  More precisely, let
$$\rho = \sum_k p_k \rho_k \in \cA$$
be an ensemble of states in a memory $\cA$.  If the symbol $k$ is not recorded,
then $\rho$ encodes all statistical information that can be extracted from the
ensemble.  But if each symbol $k$ is recorded as a state $\sigma_k$ in another
memory $\cB$, then we can let
$$\rho' = \sum_k p_k \rho_k \tensor \sigma_k \in \cA \tensor \cB.$$
If $\cB$ is abelian and the $\sigma_k$'s are distinct pure states, then the
state $\rho'$ denotes an ensemble with a record of its preparation.
The term ``ensemble'' also typically implies that the memory
$\cB$ is hidden or untransmitted.  This too is only a special
case, because memory may be hidden whether or not it is abelian.

Theorems~\ref{th:embed}, \ref{th:cap}, and \ref{th:code} together suggest that
all quantum information can be measured in the bulk limit by two numbers,
classical entropy $H$ and quantum entropy $S$.  By contrast information
capacity has more structure than information itself.  The capacity of a quantum
memory is defined by a curve that represents trade-offs between classical and
quantum entropy.  The capacity of a general quantum channel could be even more
complicated.

There are many interesting partial orderings on quantum memories besides
embeddability, bulk embeddability, and supermajorization. One natural example
is embeddability in the presence of an auxiliary memory, or \emph{stable}
embeddability.  Given memories $\cA$ and $\cB$, when is there a memory $\cC$
such that
$$\cA \tensor \cC \injects \cB \tensor \cC?$$
We do not know when $\cA$ stably embeds in $\cB$. Stable embeddability implies
bulk embeddability and is implied by embeddability, but we do not know how it
compares to supermajorization order.

\thm{th:embed} is related to a much more general question in quantum
information theory.  Let $\cE:\cA \to \cB$ and $\cF:\cC \to \cD$ be
quantum operations representing two quantum channels between general quantum
memories.  When are there operations $\cX_N$ and $\cY_N$ that make the diagram
$$\begin{array}{c@{\hspace{2cm}}c} \\[-1ex]
\rnode{a}{\cA^{\tensor N}} & \rnode{b}{\cB^{\tensor N}} \\[1.5cm]
\rnode{c}{\cC^{\tensor N(1+\eps)}} & \rnode{d}{\cD^{\tensor N(1+\eps)}}
\end{array}
\ncline[nodesep=.3]{->}{a}{b}\Aput{\cE^{\tensor N}}
\ncline[nodesep=.3]{->}{c}{d}\Aput{\cF^{\tensor N(1+\eps)}}
\ncline[nodesep=.25]{->}{a}{c}\Aput{\cX_N}
\ncline[nodesep=.25]{->}{d}{b}\Aput{\cY_N}
$$
commute with high fidelity? We can then say that the channel $\cE$ reliably
bulk-encodes in the channel $\cF$.  Theorems \ref{th:embed}, \ref{th:cap}, and
\ref{th:code} together answer the question when $\cE$ and $\cF$ are both the
identity map, with the refinement that perfect fidelity is possible when high
fidelity is possible.  In light of \thm{th:ce1}, the cross-encoding question is
also settled when $\cE$ and $\cF$ are SUCP projections.

Finally, it is well-understood that classical and quantum memory are
inequivalent resources in quantum complexity theory. For example there is a
quantum algorithm to find a collision of a 2-to-1 function with which uses
$\tO(N^{1/3})$ classical space (and $\tO(1)$ quantum space)
\cite{BHT:collision}. But if the function only has a single repeated value, the
best quantum algorithm uses $\tO(N^{1/4})$ quantum space
\cite{Heiligman:matches}.  It would be interesting to find an algorithm whose
natural space complexity is hybrid quantum memory.

\acknowledgments

The author would like to thank Daniel Gottesman, Janko Gravner, Patrick Hayden,
Dongseok Kim, Alexei Kitaev, and Bruno Nachtergaele for very helpful
discussions.  The referees were also extremely helpful.

% \bibliography{qa,qp,me,books}

\providecommand{\bysame}{\leavevmode\hbox to3em{\hrulefill}\thinspace}

\end{document}